# Mapping the Future of Particle Radiobiology in Europe: The INSPIRE Project

N.T. Henthorn[1,2, +], O. Sokol[3 +], M. Durante[3,4 *], L. De Marzi[5], F. Pouzoulet[6], J. Miszczyk[7], P. Olko[7], S. Brandenburg[8], M-J. van Goethem[8], L. Barazzuol[9,10], M. Tambas[9], J.A. Langendijk[9], M. Davídková[11], V. Vondráček[12], E. Bodenstein[13], J. Pawelke[13,14], A. Lomax[15,16], D.C. Weber[15,17,18], A. Dasu[19,20], B. Stenerlöw[20], P.R. Poulsen[21], B.S. Sørensen[21], C. Grau[21], M.K. Sitarz[21], A-C Heuskin[22], S. Lucas[22], J.W. Warmenhoven[1,2], M.J. Merchant[1,2], R.I. Mackay[23,1], K.J. Kirkby[1,2]

[1] Division of Cancer Sciences, School of Medical Sciences, Faculty of Biology, Medicine and Health, The University of Manchester, Manchester, UK

[2] The Christie NHS Foundation Trust, Manchester Academic Health Science Centre, Manchester, UK

[3] Department of Biophysics, GSI Helmholtzzentrum für Schwerionenforschung, Darmstadt, Germany

[4] Department of Condensed Matter Physics, Technische Universität Darmstadt, Darmstadt, Germany

[5] Institut Curie, PSL Research University, Radiation Oncology Department, Paris, France

[6] Institut Curie, PSL Research University, Translational Research Department, Experimental Radiotherapy Platform, Orsay, France

[7] Institute of Nuclear Physics Polish Academy of Sciences, Krakow, Poland

[8] KVI-Center for Advanced Radiation Technology, University of Groningen, Groningen, The Netherlands

[9] Department of Radiation Oncology, University of Groningen, University Medical Center Groningen, Groningen, The Netherlands

[10] Department of Biomedical Sciences of Cell and Systems, Section of Molecular Cell Biology, University of Groningen, University Medical Center Groningen, Groningen, The Netherlands

[11] Department of Radiation Dosimetry, Nuclear Physics Institute of the CAS, Prague, Czech Republic

[12] Proton Therapy Center Czech, Prague, Czech Republic

[13] OncoRay – National Center for Radiation Research in Oncology, Faculty of Medicine and University Hospital Carl Gustav Carus, Technische Universität Dresden, Helmholtz-Zentrum Dresden – Rossendorf, Dresden, Germany

[14] Helmholtz-Zentrum Dresden – Rossendorf, Institute of Radiooncology - OncoRay, Dresden, Germany

[15] Centre for Proton Therapy, Paul Scherrer Institute, Villigen, Switzerland

[16] Department of Physics, ETH, Zurich, Switzerland

[17] Department of Radiation Oncology, University of Zurich, Zurich, Switzerland

[18] Department of Radiation Oncology, University of Bern, Bern, Switzerland

[19] The Skandion Clinic, Uppsala, Sweden

[20] Department of Immunology, Genetics and Pathology, Uppsala University, Uppsala, Sweden

[21] Danish center for Particle Therapy, Aarhus University Hospital, Aarhus, Denmark

[22] Namur Research Institute for Life Sciences (NARILIS), University of Namur, Namur, Belgium

[23] Christie Medical Physics and Engineering, The Christie NHS Foundation Trust, Manchester, UK

[+] *These authors contributed equally to this work*

**\* Correspondence:**
Marco Durante
m.durante@gsi.de

**Abstract**

Particle therapy is a growing cancer treatment modality worldwide. However, there still remains a number of unanswered questions considering differences in the biological response between particles and photons. These questions, and probing of biological mechanisms in general, necessitate experimental investigation. The "Infrastructure in Proton International Research" (INSPIRE) project was created to provide an infrastructure for European research, unify research efforts on the topic of proton and ion therapy across Europe, and to facilitate the sharing of information and resources. This work highlights the radiobiological capabilities of the INSPIRE partners, providing details of physics (available particle types and energies), biology (sample preparation and post-irradiation analysis), and researcher access (the process of applying for beam time). The collection of information reported here is designed to provide researchers both in Europe and worldwide with the tools required to select the optimal center for their research needs. We also highlight areas of redundancy in capabilities and suggest areas for future investment.

**1    Introduction**

There is a growing investment in proton and heavy ion therapy worldwide, with 89 proton centers and 12 carbon centers currently in clinical operation (according to the Particle Therapy Co-Operation Group (PTCOG) [1]. Of these worldwide facilities, 31 proton centers (~35%) and four carbon centers (~33%) are located in Europe [2]. Despite the increasing adoption of particle therapy there remains a number of unanswered questions about this relatively new treatment modality [3]. These questions range widely in scope and include physical (e.g. range uncertainties or organ motion), biological (e.g. uncertainties in relative biological effectiveness and lack of clinically relevant *in vivo* data), and societal aspects (e.g. cost-effectiveness and radiotherapy demand) [4]. Many clinical centers offer beam time for research activities to address some of these questions [5]. However, access and utilization of this beam time can be difficult due to a lack of supply and/or funding. Rectifying this situation requires targeted efforts from both researchers and funders alike.

The European project "Infrastructure in Proton International Research" (INSPIRE) was created to allow researchers across Europe access to "state-of-the-art" research capabilities in centers for proton therapy. In addition, multi-ion research centers (KVI, Groningen, the Netherlands; GSI, Darmstadt, Germany) augment the particle research portfolio. INSPIRE aims to integrate research activities in protons (and heavy ions) across Europe through eight objectives:

1) Developing new infrastructure by bringing together clinical, academic, and industrial research activities.
2) Enabling access to research infrastructure for researchers in both the public and private sector.
3) Providing training for the next generation of researchers in the field.
4) Facilitating knowledge exchange to promote best research practices throughout Europe.
5) Developing joint research activities (JRAs) that will improve the facilities available within the infrastructure.
6) Developing JRAs in fields where technological challenges exist to improve European competitiveness.
7) Developing an innovation pipeline to translate research into clinical practice and industrial products.
8) To conduct research within the principles of responsible research and innovation.

The project is comprised of 17 European partners, 12 of which offer beam time through transnational access (TNA) (Table 1); a complete list of the INSPIRE partners can be found at https://protonsinspire.eu/. Further to the partners discussed in this work, the University of Namur (Belgium) is also an INSPIRE partner taking part in radiobiological research, but with their nearby partner center under development does not offer TNA through INSPIRE. However, once operational their resources will be available outside of the current INSPIRE project. Most of these partners are either clinical centers or have very close connections to clinical centers (Figure 1), for example the radiobiological capabilities of CHRISTIE and UNIMAN are shared as are those of RUG and UMCG. A close clinical link is essential to aid the design of the research at inception and to ensure its relevance and future translation to the clinic.

Further to the information hosted by each institute's website, and the information presented in this work, the following references give more information and available setups for Institut Curie [6–8], TUD [9–16], IFJ PAN [17], RUG [18–24], and GSI [25–31].

Through INSPIRE we are able to investigate important research questions together and benefit from cross-validation. An immediate example is the variability in data for proton relative biological effectiveness (RBE) that has been seen in the literature over the years [32–35]. A coordinated effort amongst the INSPIRE partners is allowing this variability to be investigated both computationally and experimentally, and the results made available to researchers across Europe through INSPIRE's experimental and modelling JRA. This systematic and coordinated approach will highlight factors leading to variation and propose mitigation strategies for future studies. These mitigation strategies will help to develop best practices for proton radiobiology research and build upon previous work on the topic [36]. Alongside coordinated research INSPIRE also seeks to improve the infrastructure available to European researchers through its TNA. Many research centers have invested significantly to develop their research, constructing accelerators, beamlines, and purchasing experimental equipment. INSPIRE also continually upgrades its research capabilities by taking research developed through JRA and making it available to the wider research community via TNA. This means that INSPIRE is able to offer the very latest technology and capabilities.

TNA provides researchers an opportunity to access beam time and funding for experiments at INSPIRE partners. The beam time is offered to all researchers and is not limited to INSPIRE partners. Furthermore, whilst the beam time is largely accessible for European researchers, up to 30% of the hours are available to researchers outside the EU. The application process is managed through the INSPIRE website (https://protonsinspire.eu). Prior to submitting the application through the online form, the researchers are advised to contact the representative of the relevant partner site to discuss the technical details of their proposed experiment. Before being transferred to an independent international user selection panel (USP), the refined application, submitted via the online form, is first assessed to ensure that the requested TNA site has the capacity and infrastructure to perform the experiment. Afterwards, the application is evaluated by at least two members of the USP for its technical and scientific excellence, as well as future potential and impact. Priority is given to users who have not had access to the TNA before. The INSPIRE website contains details about each center, links to websites, and contact information for general enquiries aimed to aid the potential researcher.

The information provided in this paper acts as a corollary to the INSPIRE website, where up-to-date information is maintained. Here, we provide details of the TNA radiobiology capabilities of each INSPIRE partner. Similar information, at least in terms of the physics capabilities, has previously been presented by the European Particle Therapy Network [37] and can be used alongside this work. Planning of a radiobiological experiment requires the knowledge of not only the beamline for the sample irradiation, but also of the available equipment and capabilities of the biological laboratories on site. The latter are essential for the sample preparation and post-processing. In this work, we aim to provide comprehensive information on the facilities available across INSPIRE. We specify details of the "physics", including location, beamlines, particle types, energies, and field sizes. We specify details of the "logistics", including details of sample types, positioning, and automation. We specify details of the "biology", including the available equipment for sample preparation and post-irradiation processing. Finally, we discuss future perspectives for ongoing development and further investment. The details provided here act as a resource for the potential researcher to select the optimal center for their experimental needs. However, it should be noted that there is often flexibility in many of the aspects we report. As such the information we provide should be used as a guide and more specific details can be obtained through communication with a specific partner or through INSPIRE's help desk. It is apparent that the capabilities, at least in terms of "physics", between many partners are similar. This level of redundancy is desirable, enabling repetition to ensure scientific rigor, however, establishing these centers requires a large investment and through INSPIRE they are able to work effectively together to ensure optimum utilization

## 2    Physics – Location, Beamlines, Particles, Energies, and Fields

A researcher often faces large heterogeneity when performing experiments between centers, with differences in protocol, setup, irradiation, and sample processing. Despite this there are a number of overlaps in beam properties and possible experiments between centers. Figure 2 shows a summary of capabilities for the INSPIRE TNA partners.

TNA providers mainly cover central and northern Europe, with a similar distribution to clinical centers (Figure 1). Geographic positioning of centers is an important factor to minimize both travel expenses and logistics. A new initiative with the South East European International Institute for Sustainable Technologies (SEEIST) [38,39] aims to enable researchers from the south east of Europe to access INSPIRE's capabilities while they are developing their own facilities.

All of the TNA providers can supply protons, with two centers, GSI and RUG, additionally offering other ion types of clinical interest, such as carbon, helium, or oxygen. As can be seen from Figure 3, in general, the energies available from the accelerator are similar between providers. The most overlapping energy region is between 120 and 190 MeV – experiments at this energy can be done at all of the partner centers. The highest possible energies can be achieved at GSI, reaching up to 1 GeV/u for heavy ions and 4.5 GeV/u for protons, with relevance to proton radiography [40] experiments, while most of the other institutes are limited to a maximum of 230-240 MeV/u. The lowest possible proton energies are offered at RUG (15 MeV) and Institut Curie (20 MeV). Energies can be further degraded before the sample to investigate increased proton linear energy transfer, with a relevance for end of range effects. Access to even lower energies can be obtained through the EU project RADIATE [41].

Eight TNA providers have a dedicated research room. This can be useful for studies that require longer irradiations and/or longer follow-up, it also gives more freedom to experiments that require a complex or non-standard sample setup. However, the cost of such studies should always be considered. Whilst the sample may be able to remain in the room post-irradiation this will often inactivate the room using valuable resources. A shared room has the downside of limited usage, due to clinical commitments, although it has the added benefit of rigorous quality assurance to a clinical standard. However, it should be noted that all partners undertake measures to ensure dosimetry and quality of beam delivery in their research rooms.

Figure 4 shows examples of beamlines for the CHRISTIE + UNIMAN, Skandion, RUG, TUD, GSI, and Institut Curie partners.

There is a range of maximum available scanned field sizes across the INSPIRE partners, shown in Figure 5. Six partners, PSI, Skandion, NPI-CAS, IFJ PAN, Au, and Institut Curie, offer the same field size ($30 \times 40$ cm$^2$). TUD and CHRISTIE + UNIMAN offer the same field size but in the landscape orientation ($40 \times 30$ cm$^2$). All partners offer a field size large enough to irradiate most *in vitro* sample types, such as tissue culture flasks or microplates. The field size may become a limitation for larger non-standard samples, or simultaneous irradiation of multiple samples. Though in some cases the field size may be increased by introducing scatterers.

Choice of reference radiation is an important aspect in general for radiobiology. The biological effects of protons are often quoted relative to the more familiar photon case, most notably the relative biological effectiveness for cell kill. A variety of reference photon qualities are used between the INSPIRE partners. Several partners have the possibility to choose between clinical LINACs and kilovoltage X-ray machines (CHRISTIE + UNIMAN, TUD, NPI-CAS, Institut Curie, RUG + UMCG), whilst the capabilities of others are more limited. The difference in reference radiation may lead to slight differences in relative effect measurements, making inter-center comparisons more complicated. However, it should be noted that this is a problem for radiobiology in general and is not limited to INSPIRE partners [42].

## 3    Logistics – Samples, Positioning, and Automation

The mode of sample irradiation is an important consideration, including sample orientation and possibility of automated handling. Monolayers of cells, grown in a flask or microplate, should not be free from media for a long duration of time to avoid drying. As such, several centers, particularly with horizontal beamlines, employ automated sample handling. Here, the sample can remain in a horizontal orientation and is lifted up only when presented to the beam for irradiation. Automated sample handling also has the added benefits of improving repeatability and minimizing access to the irradiation room, increasing sample throughput. Four centers employ automated sample handling. All the centers have the capability of a horizontal beamline, though four can additionally offer a vertical beam direction, and six offer more irradiating angles by using gantries. The sample type that can be irradiated is a limitation defined by the system. Most centers have flexibility here, with all capable of irradiating at least flasks and well plates. The sample type capability may go beyond this (as long as it can be fixed in front of the beam and meet the safety regulations of the experimental room) and should be further discussed with the partner institute. Table 2 shows a summary of these details.

Figure 6 shows examples of sample presentation to the beam at Christie + UNIMAN, RUG, Institut Curie, GSI, and AU. The system at CHRISTIE + UNIMAN (Figure 6A) employs a 6-axis robot mounted inside a hypoxia end station. The space limitations of the hypoxia cabinet mean that at most a mix of up to 36 samples can be housed at a time. The fingers of the robot are designed for T75 flasks or 96-well plates, limiting the sample type. However, other samples can be used so long as they have the same footprint as a 96-well plate or through use of customized sample holders, alternatively a large range of samples can be used without the robot. Similar to the CHRISTIE + UNIMAN system, the GSI system (Figure 6D) holds samples in the horizontal position lifting them to the beam for irradiation. This change in orientation minimizes the time that cells are free from media, ensuring a good cellular environment and avoiding sample drying. Alternatively, samples can be prepared so that the culture vessel is full of cell media, which is the case for RUG (Figure 6B) and Institut Curie (Figure 6C).

## 4    Biology – Sample Preparation and Processing

Alongside the physics capabilities, the biological equipment available at a center will often define the type and complexity of experiments that are possible. This impacts both the pre-irradiation sample preparation and post-irradiation analysis. For some experiments it is not possible to prepare samples prior to transport to the irradiating center. Similarly, it is not always possible to fix samples following irradiation ready for transport to the home institute. Table 3 gives details of the *in vitro* biological equipment available at INSPIRE partners. In most cases the equipment detailed in Table 3 is shared between the INSPIRE partner and other groups at the same institute. Therefore, these details should be used as a guide for maximum available equipment. Similarly, extra resources may be available at a partner's sister institute. Researchers requiring the use of any of this equipment should discuss their needs with the relevant partner.

Common amongst all centers is the availability of flow hoods and incubators, with TUD offering the largest capacity for sample preparation and storage. At the moment, only one center, UNIMAN, has a hypoxia station for irradiation of samples under variable oxygen tension. This offers the capability for studying the oxygen enhancement ratio and probing new fields such as the FLASH effect under strictly controlled conditions. The hypoxia station at UNIMAN is positioned directly at the beam nozzle, which prevents $O_2$ fluctuations in the sample while it is being transported from the laboratory to the irradiation facility. Additionally, the irradiation in hypoxic conditions is possible at AU and GSI, where the samples can be gassed inside specially designed containers prior the transportation to the experimental room. The availability of more sophisticated post-irradiation analysis, such as flow cytometry, FACS, mass spectrometry, PCR, and sequencing is varied amongst the partners. Similarly, the advanced microscopy available amongst the partners is varied, though the majority have fluorescent and confocal microscopes available.

While all the INSPIRE TNA partners mentioned in this work offer the environment for *in vitro* studies, the *in vivo* capabilities are slightly more limited, as seen in Figure 2. Despite the data from cell experiments being a valuable preliminary tool for studying the effects of proton beams, all of the physiological processes and their complex interplay cannot be reproduced *in vitro*, and thus the clinical treatments must first be simulated using animal models before moving onto human trials. Table 4 shows the *in vivo* capabilities of the INSPIRE TNA providers.

*In vivo* experiments bring the added complexity of ethical review. INSPIRE has a well-established ethics platform for both its TNA and JRA, which is overseen by an ethics panel comprised of international experts in the field. The partners must also follow both the official regulations of their country/state as well as those of the TNA provider. Moreover, these regulations might vary from one state to another within the same country (for example, in Germany). Ethics applications in EU generally require a FELASA (Federation of European Laboratory Animal Science Associations) certification for participating scientists that cover the duration of the relevant research. In addition to that, country-specific licenses might be required. In the latter case, exceptions can be made when the guest scientists are only irradiating the animals without leaving them at the TNA facility. The application for the ethical approval is normally done well in advance, as the review procedure can last up to several months. All of the paperwork relating to ethical approval is retained by the partner and made available to the EU upon request. In addition, for some experiments the EU requires copies of the ethical permissions prior to any experiment taking place.

## 5    Future Perspectives

As has been shown, the resources available within the INSPIRE network are state-of-the-art. Further to this a number of new centers are under development and will soon be accessible to the research community. For example, the Proteus ONE IBA center at Charleroi (Belgium) will offer both *in vivo* and *in vitro* capabilities complete with a basic *in vitro* lab and animal facility on site, with researcher access offered through partnership with Namur. Belgium is also developing a center at Leuven, which will also offer *in vitro* and *in vivo* research capabilities. Furthermore, the European project SEEIST [38,43] will develop capabilities in South-eastern Europe, filling in some geographical gaps shown in Figure 1. As well as developing a new heavy ion center the SEEIST project will have access to resources provided by INSPIRE.

There is a growing European interest into studying the effectiveness of heavy ions, with four operational carbon centers and two new centers under construction. A 2019 meeting of UK clinicians, scientists, engineers, and stakeholders began the process of considering future UK development of heavy ion therapy. There are also ongoing investigations into the clinical utilization of other particle types. For example, Helium has been seen as an intermediate between protons and carbon [44–46]. Other studies investigate the possibilities of combining multiple beams within one treatment plan to ensure a more uniform RBE distribution [47], or better treatment of hypoxic tumors [48]. The INSPIRE network is well placed for the associated radiobiological investigations here, in particular with the partner institutes GSI and RUG.

There has been a worldwide renewed interest in radiotherapy delivery techniques and improved normal tissue sparing. For example, spatially fractionated proton therapy [49–52] and ultra-high dose rate (FLASH) [53–56]. In these cases, the radiobiological mechanism driving the effect remains elusive. In particular, the differences between photon and particle therapy requires further investigation. Alongside this, the combination of particle therapy with immunotherapy [57,58] is an exciting treatment that requires mechanistic understanding. Again, the INSPIRE network provides resources for investigation here, particularly through *in vivo* work, with results being directly useful for clinical adoption.

   *In vivo* radiobiological research is a crucial step along the path to clinical implementation. Eight of the 12 partners discussed in this work are currently performing *in vivo* research. Further to this, CHRISTIE + UNIMAN are beginning development of a second beamline for *in vivo* work. Skandion are also in the early stages of planning future *in vivo* work. This added capacity, and the currently available capacity, is sure to aid in the clinical efficacy of proton therapy.

   The connection between research activities and clinically relevant questions must be made stronger. There are close links between many INSPIRE partners and clinical centers, which aids in this connection. However, it is important that the clinical community become more involved with research at inception. With a limited amount of finances this will ensure prioritization of the most pertinent research and advance clinical translation, all for the benefit of the patient.

## 6   Conclusion

   In this work we have given details about the radiobiological capabilities of partners involved in the INSPIRE project, including how the resources can be accessed. It is clear that whilst there are a number of differences between the partners there are also a number of similarities. This allows for investigations into the cause of variance in published radiobiological data, such as the planned joint experiment of the INSPIRE partners. However, establishing these research centers requires significant investment and, as can be seen, many of the capabilities are already in place. More effort must be made to develop and utilize the resources currently available to us. Efforts are being made to further increase *in vivo* capabilities, whilst *in vitro* research is invaluable for identifying and probing mechanisms, *in vivo* research is crucial for clinical adoption. Also required here is a closer relationship with clinical partners, ensuring a good direction for future research. With a renewed interest in radiotherapy delivery techniques, and the unknown biological mechanisms, now is certainly and exciting time for particle radiobiology. Mechanisms that the INSPIRE network is well placed to address.

## 7   Conflict of Interest

   The authors declare that the research was conducted in the absence of any commercial or financial relationships that could be construed as a potential conflict of interest.

## 8   Author Contributions

   MD designed the structure of the manuscript. NTH and OS wrote the manuscript with input from the other authors. OS and MD provided information for GSI. LDM and FP provided information for Institut Curie. JM and PO provided information for IFJ PAN. SB, M-JvG, and LB provided information for RUG. MT and JAL provided information for UMCG. MD and VV provided information for NPI-CAS. EB and JP provided information for TUD. AL and DSW provided information for PSI. AD and BS provided information for Skandion. PRP, BSS, CG, and MKS provided information for AU. A-CH and SL provided information for Namur. NTH, JWW, MJM, RIM, and KJK provided information for UNIMAN and CHRISTIE. MD leads the radiobiology work package of the INSPIRE project. KJK leads the INSPIRE project. All authors reviewed and agreed the manuscript.

## 9   Funding

This work was funded by the European Union's Horizon 2020 research and innovation programme under grant agreement No 730983 (INSPIRE).

**Tables:**

| Centre | Abbreviation | Location | Website |
|---|---|---|---|
| Aarhus University | AU | Aarhus, Denmark | https://www.en.auh.dk/departments/the-danish-centre-for-particle-therapy/ |
| The Christie NHS Foundation Trust | CHRISTIE | Manchester, UK | https://www.christie.nhs.uk |
| GSI Helmholtz Centre for Heavy in Research | GSI | Darmstadt, Germany | https://www.gsi.de/work/forschung/biophysik.htm |
| The Henryk Niewodniczański Institute of Nuclear Physics Polish Academy of Sciences | IFJ PAN | Kraków, Poland | https://inspire.ifj.edu.pl/en/index.php/dostep-do-infrastruktury-badawczej/ |
| Curie Institute | Institut Curie | Paris, France | https://institut-curie.org/page/research-and-development-proton-therapy-center |
| Nuclear Physics Institute of the Czech Academy of Sciences | NPI-CAS | Prague, Czech Republic | http://www.ujf.cas.cz/en/ |
| Paul Scherrer Institute | PSI | Zurich, Switzerland | https://www.psi.ch/en |
| University of Groningen | RUG | Groningen, Netherlands | https://www.rug.nl/kvi-cart/research/facilities/agor/ |
| Skandion Clinic | Skandion | Uppsala, Sweden | https://skandionkliniken.se/ |
| Technical University of Dresden | TUD | Dresden, Germany | https://www.oncoray.de/research/offer-for-users/ |
| University Medical Center Groningen | UMCG | Groningen, Netherlands | https://www.umcgradiotherapie.nl/en/umcg-groningen-department-of-radiation-oncology |
| University of Manchester | UNIMAN | Manchester, UK | https://www.bmh.manchester.ac.uk/research/domains/cancer/proton/ |

**Table 1:** The INSPIRE partners offering equipment and support for radiobiological experiments through transnational access.

| Centre | Beam direction (H/V) | Gantry | Automated sample exchange | Sample type | | | |
|---|---|---|---|---|---|---|---|
| | | | | Flask | Petri Dish | Well Plate | Other Vessels |
| AU | H | x | x | ✓ | x | ✓ | ✓ |
| GSI | H | x | ✓ | ✓ | x | ✓ | ✓ |
| IFJ PAN | H | ✓ | x | ✓ | ✓ | ✓ | ✓ |
| Institut Curie | H + V | ✓ | ✓ | ✓ | ✓ | ✓ | x |
| NPI-CAS | H + V | ✓ | x | ✓ | ✓ | ✓ | x |
| PSI | H + V | ✓ | x | ✓ | ✓ | ✓ | ✓ |
| RUG + UMCG | H | x | ✓ | ✓ | ✓ | ✓ | ✓ |
| Skandion | H + V | ✓ | x | ✓ | ✓ | ✓ | ✓ |
| TUD | H | x | x | ✓ | ✓ | ✓ | ✓ |
| CHRISTIE + UNIMAN | H | ✓ | ✓ | ✓ | x | ✓ | ✓ |

**Table 2:** Beamline and radiobiological sample details of the INSPIRE partners. All centers can offer a horizontal beamline, with the four able to irradiate samples from above or at user-defined angles using a gantry. There is flexibility in sample types, but the majority of centers have the ability to irradiate flasks and well plates.

| Centre | # of Laminar Flow Cabinets | # of Incubators | Hypoxia Irradiation Station | Chemical Hood | Flow Cytometry | FACS | Biological Mass Spectrometry | PCR | Sequencing | Fluorescent Microscope | Confocal Microscope | Super Resolution Microscope |
|---|---|---|---|---|---|---|---|---|---|---|---|---|
| AU | 1 | 1 | x | x | x | x | x | ✓ | x | x | ✓ | x |
| GSI | 2 | 4 | x | ✓ | ✓ | ✓ | x | ✓ | x | ✓ | ✓ | x |
| IFJ PAN | 2 | 1 | x | ✓ | x | x | x | ✓ | x | ✓ | ✓ | x |
| Institut Curie | 1 | 1 | x | ✓ | ✓ | ✓ | ✓ | ✓ | ✓ | ✓ | ✓ | ✓ |
| NPI-CAS | 2 | 3 | x | ✓ | x | x | x | ✓ | ✓ | ✓ | x | x |
| PSI * | 0 | 1 | x | x | x | x | x | x | x | x | x | x |
| RUG + UMCG | 2 | 2 | x | ✓ | x | x | x | x | x | x | ✓ | x |
| Skandion | 4 | 4 | x | ✓ | ✓ | ✓ | ✓ | ✓ | ✓ | ✓ | ✓ | ✓ |
| TUD | 6 | 12 | x | ✓ | ✓ | x | x | ✓ | x | ✓ | ✓ | x |
| CHRISTIE + UNIMAN | 5 | 5 | ✓ | ✓ | ✓ | ✓ | ✓ | ✓ | ✓ | ✓ | ✓ | ✓ |

**Table 3:** *In vitro* biological analysis equipment available at the INSPIRE partners. *Biological equipment at PSI is available at a partner institute and will need to be discussed.

| Centre | Animals | Capacity (max. No. of animals) | Max. days before irradiation | Onsite Immobilization | Onsite Anesthesia | Models used | Imaging | Histology |
|---|---|---|---|---|---|---|---|---|
| AU | Rats, Mice | 80 Rats, 200 Mice | 7 | x | x | Normal tissue and a range of tumor models (syngenic and xenografts) | x | x |
| GSI | Rats, Mice | 80 | 7 | x | x | x | x | ✓ |
| IFJ PAN | Rats, Mice, Hamsters | 100 | 7 | ✓ | ✓ | x | MRI | x |
| Institut Curie | Rats, Mice | 100 Rats, 40 Mice | A few months | ✓ | ✓ | Normal tissue and a range of tumor models (syngenic and xenografts), orthotopic grafts, specific tissue toxicity assays | CT, X-ray, OCT, Bioluminescence | ✓ |
| PSI * | Mice, Zebrafish | - | - | ✓ | ✓ | - | - | - |
| RUG + UMCG | Rats, Mice | 132 Rats, 264 Mice | 7 | ✓ | ✓ | Normal tissue | ✓ | ✓ |
| TUD | Rats, Mice, Zebrafish | 100 | 7 | ✓ | ✓ | Zebrafish embryo strain wild type AB; NMRI nu/nu Nude, C57Bl/6JRj and C3H/HeNRj | CT, X-ray, MRI, Proton radiography, Bioluminescence, PET, Ultrasound | ✓ |

**Table 4:** *In vivo* capabilities available at the INSPIRE partners. *In Vivo* irradiation at PSI has previously been done, but capacities and equipment need to be discussed.

**Figure Captions:**

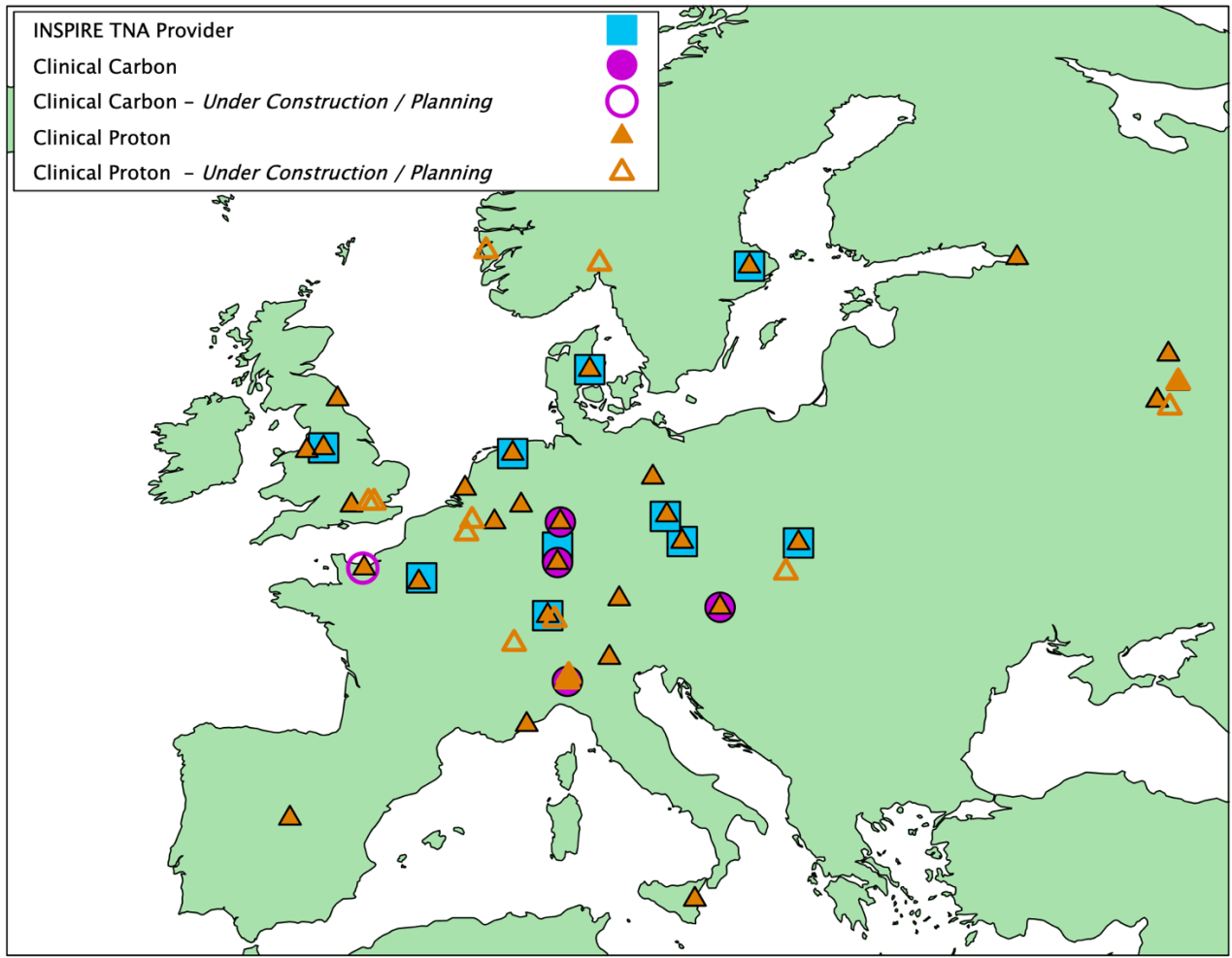

**Figure 1:** European clinical proton therapy centers (closed triangle, 26 centers), carbon therapy centers (closed circle, 4 centers), and INSPIRE partners offering radiobiological TNA (closed squares, 12 centers – there is some overlap between centers). Open symbols show centers currently in the planning stage or under construction. Information is from the PTCOG website [1].

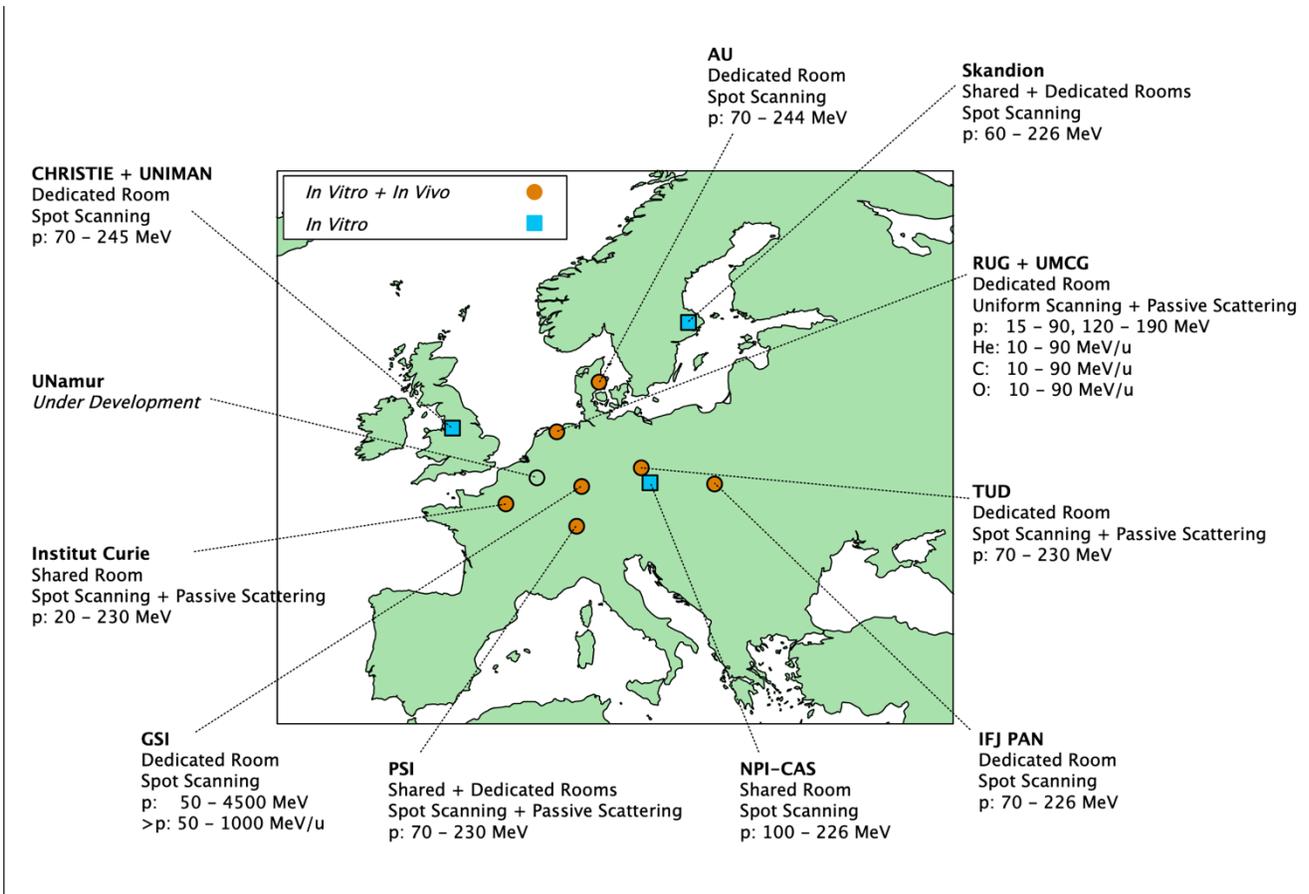

**Figure 2:** INSPIRE partners offering radiobiological investigation with particles. The quoted energies are as extracted from the beamlines, lower energies are available with beam degraders. Centers offering both *in vitro* and *in vivo* experiments are marked with orange circles, while those offering only *in vitro* experiments are shown as blue squares.

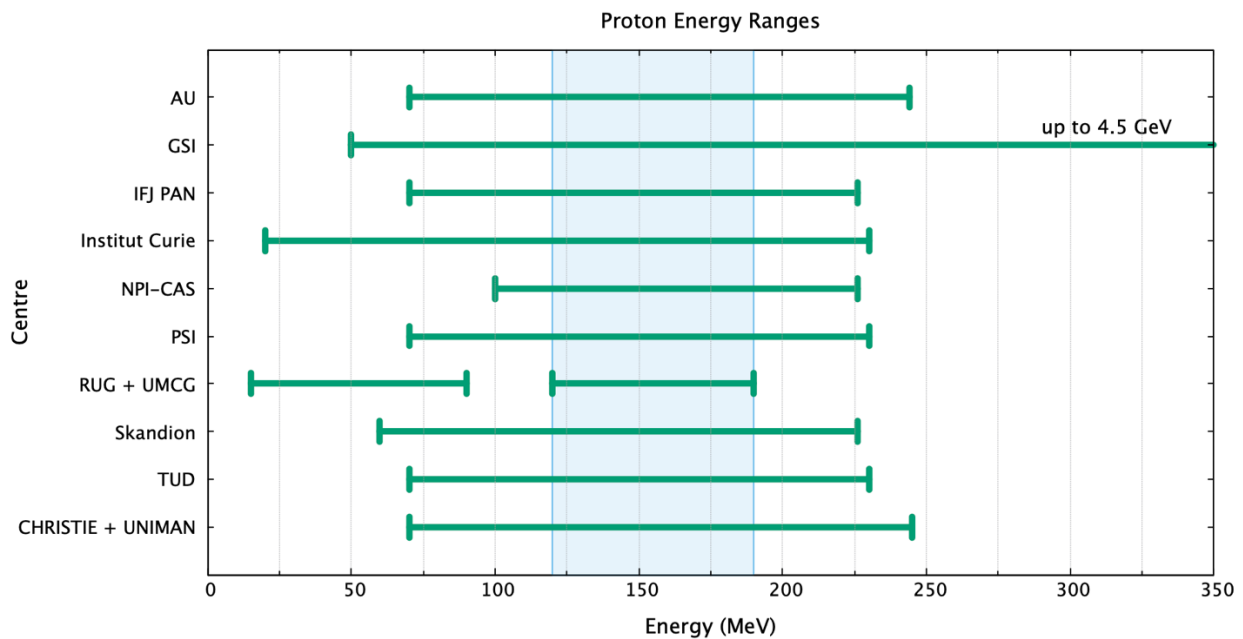

**Figure 3:** Proton energies available at INSPIRE partners as extracted from the accelerator. The highest energy is available at GSI (up to 4.5 GeV). The lowest energies are available at RUG (15 MeV) and Institut Curie (20 MeV). The overlapping region (shaded area) is between 120-190 MeV. Energies can be further degraded in front of the sample.

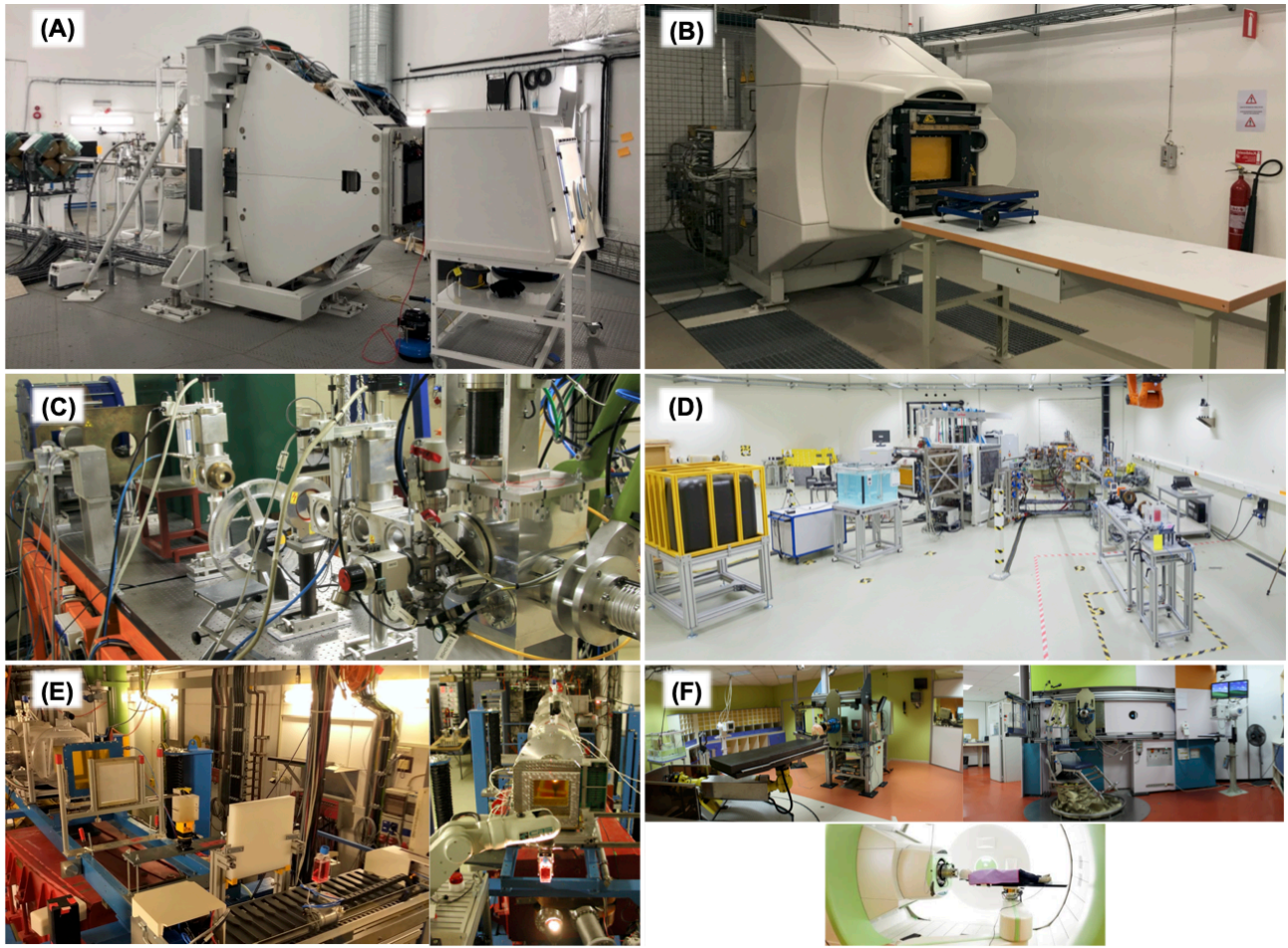

**Figure 4:** Beamline setup for **(A)** UNIMAN + CHRISTIE, **(B)** Skandion, **(C)** RUG, **(D)** TUD, **(E)** GSI, and **(F)** Institut Curie. UNIMAN has a beamline leading to a Varian scanning nozzle, samples are placed in front of the nozzle (pictured is a hypoxia cabinet). Skandion has a beamline leading to an IBA scanning nozzle, samples are placed on an adjustable table in front of the nozzle. RUG has a flexible beamline setup using optical benches; picture shows a study on the effect of magnetic fields in combination to proton irradiation [23]. TUD has two beamlines in the dedicated experimental room, one with a pencil beam scanning nozzle (left) and one static beamline (right). In the picture, setups with water tank and beam dump at the scanning beamline and passive double scattering setup for radiobiological experiments at the static beamline are shown. GSI shows the beamline setup for "Cave A", equipped with the robotic system for sample exchange. Institut Curie shows three irradiation rooms; "Room Y1" – horizontal beam up to 201 MeV (left), "Room Y2" – horizontal beam up to 76 MeV (right), and "IBA Room" – gantry up to 230 MeV (bottom).

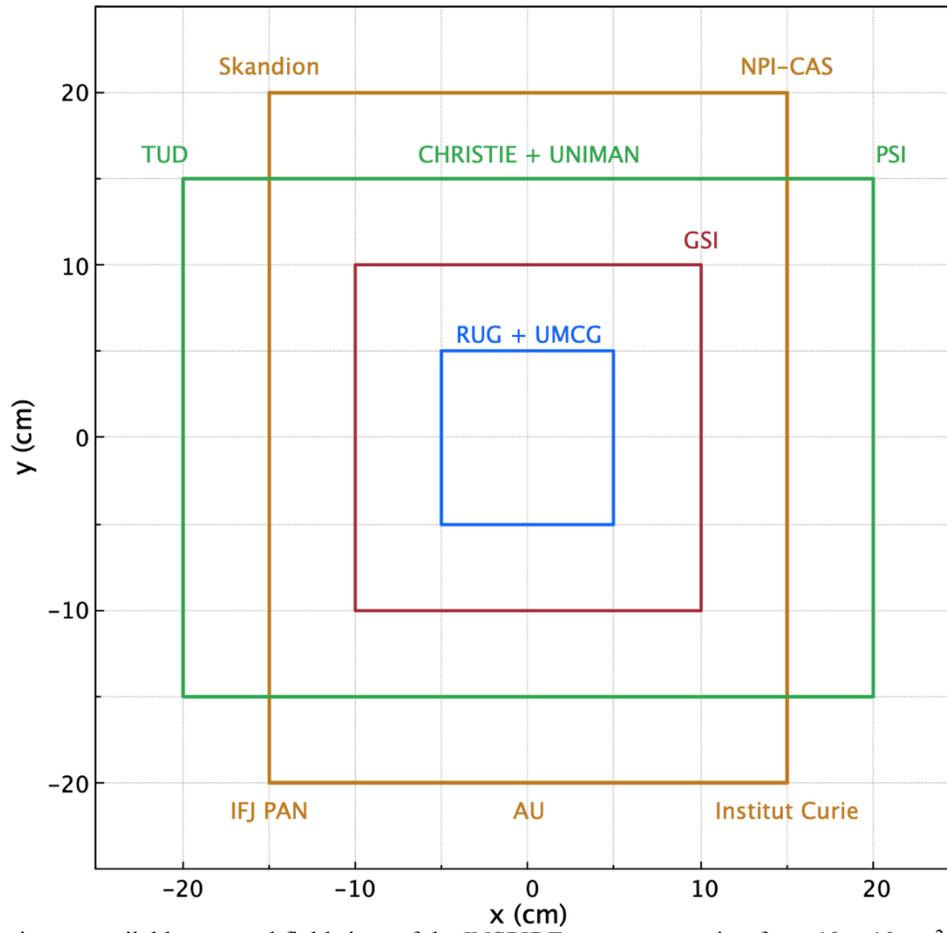

**Figure 5:** Maximum available scanned field sizes of the INSPIRE partners, ranging from $10 \times 10$ cm² to $30 \times 40$ cm². Larger field sizes may be available by introducing scatterers.

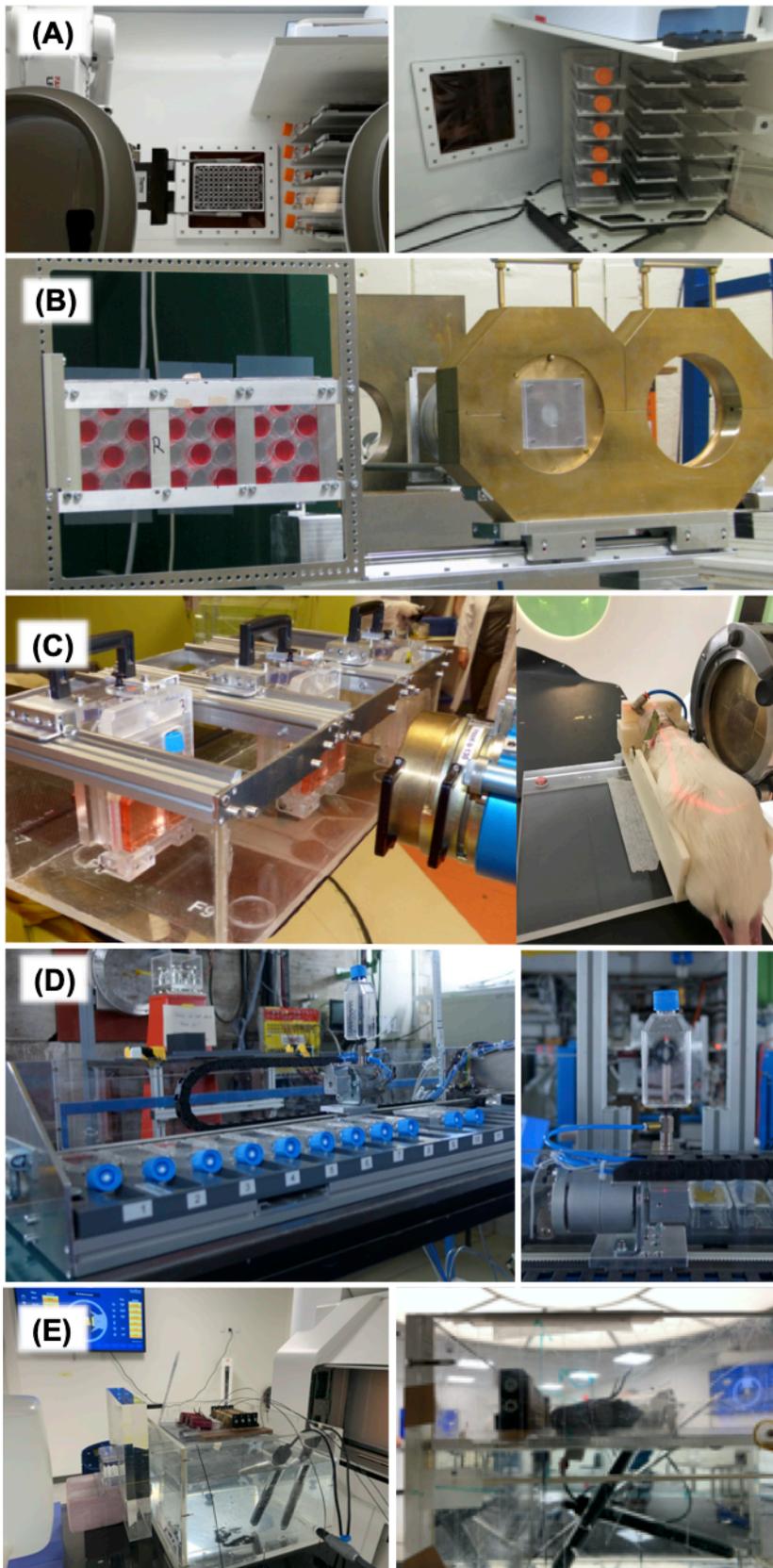

**Figure 6:** Setup for sample irradiation at **(A)** CHRISTIE + UNIMAN, **(B)** RUG, **(C)** Institut Curie, **(D)** GSI, and **(E)** AU. The CHRISTIE + UNIMAN system is a 6-axis robotic arm mounted in a hypoxia cabinet, allowing irradiation at different oxygen tensions from 0.1-20%. The robot picks samples from a "hotel" and holds them in front of a beam window within the cabinet, before either replacing the sample to the hotel or moving to an automated fixation system (left). The hotel can house up to 36 samples, a mix of T75

flasks or 96-well plates (right). The RUG system shows the sequential irradiation of three 12-well plates. Wells are filled with cell media and sealed with parafilm. The Institut Curie system shows sequential irradiation of six *in vitro* samples (left), and immobilized *in vivo* irradiation (right). The GSI system allows for sequential irradiation of 16 tissue culture flasks. The flasks remain in the horizontal position whilst not being irradiated (left), preventing the cell layer inside from drying. The robotic system lifts the sample and presents it to the beam (right), replacing it when irradiation is complete. The AU system shows an *in vivo* setup for mouse leg irradiation.